\documentclass[aip,reprint]{revtex4-1}
\usepackage[english]{babel}
\usepackage{graphicx}
\usepackage{dcolumn}
\usepackage{bm}
\usepackage[latin1]{inputenc}
\usepackage{graphicx}
\usepackage{amsmath}
\usepackage{amssymb}
\usepackage{pifont}
\usepackage{epstopdf}
\usepackage{color}

\renewcommand{\Re}{\mathrm{Re}}

\begin{document}

\title{Instability of particle inertial migration in shear flow}
\author{Evgeny S. Asmolov}
\affiliation{Frumkin Institute of Physical Chemistry and
Electrochemistry, Russian Academy of Science, 31 Leninsky Prospect,
119071 Moscow, Russia}
\affiliation{Institute of Mechanics, Lomonosov Moscow State
University, 119991 Moscow, Russia}

\author{Tatiana V. Nizkaya}
\affiliation{Frumkin Institute of Physical Chemistry and
   Electrochemistry, Russian Academy of Science, 31 Leninsky Prospect,
   119071 Moscow, Russia}

\author{Jens Harting}
\affiliation{Helmholtz Institute Erlangen-N\"urnberg for Renewable Energy,
Forschungszentrum J\"ulich,\\ Cauerstrasse 1, 91058 Erlangen,
Germany}

\affiliation{Department of Chemical and Biological Engineering and Department
of Physics, Friedrich-Alexander-Universit{\"a}t Erlangen-N{\"u}rnberg,
Cauerstrasse 1, 91058 Erlangen, Germany}

\author{Olga I. Vinogradova}
\email[Corresponding author: ]{oivinograd@yahoo.com}
\affiliation{Frumkin Institute of Physical Chemistry and
   Electrochemistry, Russian Academy of Science, 31 Leninsky Prospect,
   119071 Moscow, Russia}

\date{\today }

\begin{abstract}

In a shear flow particles migrate to their equilibrium positions in the microchannel. Here we demonstrate theoretically that if particles are inertial, this equilibrium can become unstable due to the Saffman lift force. We derive an expression for the critical Stokes number that determines the onset of instable equilibrium.   We also present results of lattice Boltzmann simulations for spherical particles and prolate spheroids to validate
the analysis. Our work provides a simple explanation of several unusual phenomena observed in earlier experiments and computer simulations, but never interpreted before in terms of the instable equilibrium.

\end{abstract}

\maketitle

\section{Introduction}

In shear flows particles experience an inertial lift force which induces their migration across streamlines.
This effect was first discovered for
neutrally buoyant particles in tubes\cite{segre1962behaviour} and is currently widely
employed to separate particles in microfluidic devices\cite{stoecklein2018nonlinear}.
% The inertial lift force results from the non-linear interaction between the perturbation flow induced by the particle and the background flow.
In an unbounded shear flow, the Saffman lift force\cite%
{saffman1965lift} emerges when a  slip velocity (i.e. a difference between
particle velocity and fluid velocity at the particle center) becomes finite. The
disturbance of the flow in this case is caused by a streamwise drag force
and momentum released into the fluid. In the channel (wall-bounded) flows, another type of the
lift force emerges for a freely rotating and translating particle. Namely, a
neutrally buoyant lift force \cite{Ho:Leal74,Vass:Cox76}, which is due to the
curvature of the undisturbed velocity profile or wall effects.

The inertial migration of particles is traditionally considered as a
quasi-steady process. This implies that the particle inertia is neglected, so that hydrodynamic forces (the drag, the
neutrally buoyant lift, the Dean force \cite{di2009inertial}) as well as any
external forces are balanced, i.e. $\mathbf{F}(\mathbf{x}%
_{p}\mathbf{,V,U})=0$, where $\mathbf{x}_{p}$ and $\mathbf{V}$ are
the particle position and velocity, correspondingly, and $\mathbf{U}$ is the fluid velocity. Besides, it is commonly considered that the Saffman lift emerges only  under forces acting in the streamwise direction (e.g. non-neutrally buoyant particles under gravity in vertical
channels). Such a quasi-steady approach allows one to infer the particle velocity $\mathbf{V}(\mathbf{x})$ by using the lift and drag
coefficients. Since the drag coefficient is positive-definite, the behavior
of the particle is controlled by the variation of the lift force across the
channel: the zeros of the lift force correspond to the particle equilibrium
positions and its gradients define their stability.
In computer simulations, the lift force on a particle at different positions
can be measured independently and is often used to predict the particle
behavior\cite{shi2020lift}. By contrast,  direct experimental
measurements of the lift force are impossible. This force is usually calculated  from the
measured migration velocity%
\cite{hood2016direct}.

The particle inertia is characterized by the Stokes number $\mathrm{St}=2\rho_{p}\Re_{G}/\left(
9\rho \right)$, where $\rho _{p}$ and $\rho$ are the particle and
fluid densities, $\Re _{G}=Ga^{2}/\nu$ is the particle Reynolds number defined using
the particle radius $a$, shear rate $G$ and kinematic
viscosity of the fluid $\nu$. Clearly, at sufficiently large $\Re_{G}$ the Stokes number can become finite even for neutrally buoyant
particles with $\rho _{p}/\rho =1$. Consequently, when such  particles migrate  across the  streamlines, they accelerate by the fluid. In this case, the momentum exchange between the fluid and the particle generates the Saffman lift force, and the quasy-steady  approach is no longer applicable.

Despite this obvious fact, the correctness of the quasi-steady approach at  $\Re _{G}>1$  is still not under dispute,
and migration phenomena
are commonly analyzed in terms of the dependence of the  lift force on the particle position,  its zeros and their bifurcations\cite%
{shi2020lift,fox_schneider_khair_2021}.
As one example, numerous experiments with
neutrally buoyant particles in circular tubes found that at high channel
Reynolds numbers the Segre-Silberberg equilibrium position shifts towards
the wall, but some particles migrate towards the center to form an inner annulus
\cite{matas2003inertial,morita2017equilibrium,nakayama2019three}. However, their interpretation implies that in the long
run all particles will focus at the zeros of the lift curve, although note that there have been some suggestions that the inner annulus is a second ``true'' equilibrium
position\cite{matas2003inertial,nakayama2019three} or represents only a
transient configuration \cite{morita2017equilibrium}.
Another example refers to the computer simulations of the inertial behavior of spheroids in shear flow that is currently a subject of active research \cite%
{qi2003rotational,huang2012rotation,rosen2015dynamical,rosen2016quantitative}%
. It is well known that at large $\Re_{G}$ and $\mathrm{St}$ prolate
spheroids undergo a series of transitions between different rotational
regimes. All these studies assume that spheroids move with the velocity
of the fluid that is equivalent to a decoupling between rotational and translational
motion. This, in turn, implies that the equilibrium is always
stable, which is by no means obvious at large $\mathrm{St}$.

In the present paper, we analyse the equilibrium
state of torque- and force-free spherical particles in the shear flow. We show that at finite $\mathrm{St}$ this equilibrium becomes unstable due to the Saffman lift force. Using lattice Boltzmann simulations we then verify our theoretical predictions for spheres and
prolate spheroids.

Our paper is organized as follows. In Sec.~\ref{s2} we derive equations of the particle
motion in an unbounded shear flow and obtain a stability criterion for small $\Re_{G}$ and finite $\mathrm{St}$. In Sec.~\ref{migr_norm} we show how
particle inertia and Saffman lift modify the particle migration velocity
under a transverse force in wall-bounded flows. The lattice Boltzmann method is described in Sec.~\ref{s4}.
In Sec.~\ref{res} we present the simulation results that validate our theoretical
predictions for heavy spheres and prolate spheroids at
finite $\Re_{G}$. Our conclusions are summarized in Sec.~\ref{con}.
Appendix A generalizes the Saffman formula  to the case of a particle translating in a transverse direction with a constant velocity and accelerating in a streamwise direction.

\section{Stability of particle motion in an unbounded shear flow}

\label{s2}

\begin{figure}[tbp]
\centering
\includegraphics[width=1\columnwidth]{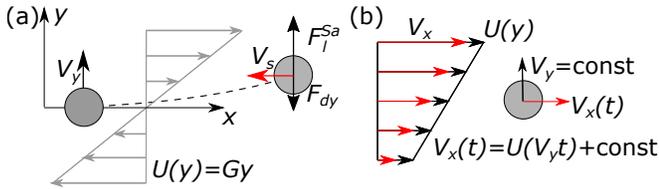}
\caption{(a) Sketch of the particle motion in a shear flow. A particle
translating across the streamlines experiences a transverse drag $F_{dy}$
and a lift force $F_{l}^{Sa}$. (b) Particle velocity in a neutral equilibrium. }
\label{fig:sketch}
\end{figure}

We begin with the migration of a force- and torque-free spherical
particle of radius $a$ and density $\rho _{p}$ in an unbounded linear shear
flow, $\mathbf{U^{\prime }}=U^{\prime }\mathbf{e}_{x},$ $U^{\prime }=Gy$
where $\mathbf{e}_{x}$ is the unit vector along the $x-$axis (see Fig.~\ref%
{fig:sketch} (a)). The particle Reynolds number $\Re_{G}=\rho Ga^{2}/\mu$ is
finite, where $\rho $ and $\mu $ are the fluid density and viscosity. The
initial particle position is $\mathbf{x}_{p0}^{\prime }=\left(
x_{p0}^{\prime },y_{p0}^{\prime },z_{p0}^{\prime }\right) =\mathbf{0}$.

The equations of particle motion in dimensional variables read

\begin{eqnarray}
m\dfrac{d\mathbf{V}^{\prime }}{dt^{\prime }} &=&\mathbf{F}^{\prime }\mathbf{,%
}  \label{motion} \\
\dfrac{d\mathbf{x}_{p}^{^{\prime }}}{dt^{\prime }} &=&\mathbf{V}^{\prime },
\label{motion1}
\end{eqnarray}%
where $m=4/3\pi \rho _{p}a^{3}$ is the particle mass, $\mathbf{V}^{\prime
}=\left( V_{x}^{\prime },V_{y}^{\prime }\right) $ is the particle
translational velocity, and $\mathbf{F}^{\prime }=\left( F_{x}^{\prime
},F_{y}^{\prime }\right) =\mathbf{F}_{d}^{\prime }+\mathbf{F}_{t}^{\prime }+%
\mathbf{F}_{l}^{\prime Sa}$ is the hydrodynamic force. Here, $\mathbf{F}%
_{d}^{\prime }$ is the quasi-steady drag force and $\mathbf{F}_{t}^{\prime }$
is an unsteady force due to particle acceleration which includes the Basset
and added-mass forces\cite{Maxey1983}. The Saffman lift force $\mathbf{F}%
_{l}^{\prime Sa}=F_{l}^{\prime Sa}\mathbf{e}_{y}$ is proportional\textbf{\ }%
to the momentum released by the particle into shear flow.
%In our flow it is positive when the particle pushed by the fluid in positive $x-$direction an negative in the opposite case, see Section \ref{sec:lift} for details.
Note that we do not include the equation for the rotational velocity since it has little effect on the sphere's dynamics in shear flows\cite%
{bagchi2002}.

A steady-state solution of Eqs.(\ref{motion}) and (\ref{motion1}) in the absence of
external force is
\begin{equation}
\mathbf{V}_{0}^{\prime }=\mathbf{0,\quad x}_{p0}^{\prime }=\mathbf{0.}
\end{equation}%

To distinguish between the stable and unstable equilibrium states we employ the linear stability analysis. Let us consider how the particle at the equilibrium position reacts to a small disturbance in the initial velocity. For small (slip Reynolds number) $%
\Re _{V}=a\left\vert \mathbf{V^{\prime }-U^{\prime }}\left( y_{p}\right)
\right\vert /\nu $ Eqs.(\ref{motion}) and (\ref{motion1}) can be linearized by expanding $\mathbf{V^{\prime }}$.
The drag force then takes the form $%
F_{dx}^{\prime }=-6\pi \mu af_{x}(V_{x}^{\prime }-Gy_{p}^{\prime })$, $%
F_{dy}^{\prime }=-6\pi \mu af_{y}V_{y}^{\prime }$, where $f_{x}\left( \Re
_{G}\right) $ and $f_{y}\left( \Re _{G}\right) $ are the correction factors
accounting for the effect of fluid inertia at finite $\Re _{G}$. At small $%
\Re _{G}$ they converge to unity.

It is now convenient to introduce dimensionless variables by scaling the velocities by $Ga,$ the
coordinates by $a,$ the time by $G^{-1},$ and the forces by $6\pi \mu
a^{2}G. $ The linearized equations can then be formulated as
\begin{eqnarray}
\mathrm{St}\dfrac{dV_{x}}{dt} &=&-f_{x}\left( V_{x}-y_{p}\right) +F_{tx}%
\mathbf{,}  \label{mot_dl1} \\
\mathrm{St}\dfrac{dV_{y}}{dt} &=&-f_{y}V_{y}+F_{l}^{Sa}+F_{ty},
\label{mot_dl2} \\
\dfrac{dy_{p}}{dt} &=&V_{y},  \label{mot_dl3}
\end{eqnarray}%
where $\mathrm{St}$ is the Stokes number,
\begin{equation}
\mathrm{St} =\frac{2\rho _{p}Ga^{2}}{9\mu }=\frac{2\rho _{p}}{9\rho }\Re
_{G}.  \label{st}
\end{equation}

The origin of instability can be understood as follows. Assume a small disturbance of the particle from its equilibrium position leading to a positive transverse velocity $V_{y}$ (see Fig.~\ref{fig:sketch}(a)). The emerging
negative transverse drag $F_{dy}=-f_{y}V_{y}<0$ should  tend to stop the particle.
However, since   the particle enters the region of a larger fluid velocity, it
lags behind the fluid due to its inertia. The particle that is subject to a drag force, $F_{dx}=-f_{x}V_{s}>0,$ where
$V_{s}=V_{x}-y_{p}<0$ is the slip velocity,  begins to accelerate in the
streamwise direction. This, in turn, should induce a positive Saffman lift force $%
F_{l}^{Sa}>0$ which may exceed the transverse drag by leading to further
acceleration of the particle in the transverse direction. We refer this situation to as unstable.

By contrast, when the equilibrium is neutral, which implies that small disturbances neither grows nor disappear,
$V_{y}$ and
$V_{s}$ are constant (see Fig.~\ref{fig:sketch}(b)). The acceleration of a particle is
also constant and equal to the fluid acceleration along its
trajectory
\begin{equation}
\frac{dV_{x}}{dt}=V_{y}\dfrac{dU}{dy}\mathbf{e}_{x}=V_{y}\mathbf{e}_{x}=%
\mathrm{const.}  \label{dvt}
\end{equation}%

It follows from Eq.~(\ref{mot_dl2}) that if $dV_{y}/dt=0$, the unsteady force $F_{ty}$ and the inertia term
in the left-hand side vanish. The condition of a neutral equilibrium can then be formulated as
\begin{equation}
F_{l}^{Sa}-f_{y}V_{y}=0.  \label{crit}
\end{equation}

To apply the criterion (\ref{crit}) it is necessary to calculate the lift force $%
F_{l}^{Sa}$ for the neutral equilibrium, which is not straightforward.
In the classical Saffman theory~\cite{saffman1965lift} the lift force is
proportional to (constant) $V_{s} $. In our case the situation is different since
the particle accelerates in the streamwise direction, and the lift force is related to the particle acceleration given by Eq.\eqref{dvdt} (see
Appendix A for a derivation) as
\begin{equation}
F_{l}^{Sa}=C_{l}^{Sa}\mathrm{St}V_{y}\quad \text{as}\quad \Re _{V}\ll \Re
_{G}\ll 1.  \label{saffman_vy}
\end{equation}%
Here $C_{l}^{Sa}=0.343\Re _{G}^{1/2}$ is the Saffman lift
coefficient, which characterizes the lift-to-drag
ratio. One can expect that the lift force at finite $\Re _{G}$ is described by \eqref{saffman_vy}, i.e.
\begin{equation}
F_{l}^{Sa}=C_{l} \mathrm{St}V_{y}\quad \text{as}\quad
\Re _{V}\ll 1,  \label{saffman_re}
\end{equation}%
but $C_l \neq C_{l}^{Sa}$ and its dependence on $ \Re _{G}$ has to be calculated.

By balancing the lift (Eq.\eqref{saffman_re}) and the transverse drag (%
Eq.\eqref{mot_dl2}) forces, we can find the critical Stokes number
\begin{equation}
\mathrm{St}_{cr}=\dfrac{f_{y}}{C_{l}},  \label{st_cr}
\end{equation}%
which determines the onset of the unstable equilibrium. For $\mathrm{St}<\mathrm{St}_{cr}$ the equilibrium is stable, but when $\mathrm{St}>\mathrm{St}_{cr}$, the lift force becomes larger than the transverse drag, $F_{l}^{Sa}>f_{y}V_{y}$, and the unstable regime develops. Note that at small $\Re _{G}$ the value of $C_{l}$ can be found using (\ref{lift}), and the correction factor is $f_{y}=1.$ Therefore, in this limiting case $\mathrm{St}_{cr}=2.92\Re
_{G}^{-1/2}$.

Eq.(\ref{st}) can be used to reformulate Eq.~(\ref{st_cr}) as
\begin{equation}
\left[ \dfrac{9f_{y} }{2\Re _{G}C_{l} }\right] _{cr}=\frac{\rho _{p}}{\rho }.  \label{impl}
\end{equation}%
This equation can be seen as an implicit condition on the critical Reynolds number. Thus, when the Saffman theory is valid, using Eq.~(\ref{lift}) one can obtain from (\ref{impl})
that
\begin{equation}
\Re _{cr}=5.56\left( \frac{\rho_p }{\rho}\right) ^{-2/3}\ll 1\text{\quad
as\quad }\rho_p /\rho \gg 1.  \label{re_cr}
\end{equation}%
$\allowbreak $
It indicates that at a
large density ratio, e.g. for aerosol particles the critical Reynolds numbers is small.
Say, for water droplets in air ($\rho _{p}/\rho \simeq 800$)
the instability is expected at $\Re _{G}>\Re _{cr} \simeq 0.065.$

For smaller density ratios, including $\rho _{p}/\rho =1$
(neutrally buoyant particles), the instability should also occur, but at finite
$\Re _{G}$.
To generalize the scaling equation (\ref{re_cr}) to the case of
finite $\Re _{G}$ we have to calculate $f_{y}\left( \Re
_{G}\right) $ and $C_{l}\left( \Re _{G}\right)$.
Moreover, in practice we normally deal with a wall-bounded flow, termed the Couette flow, where these
coefficients depend on the channel thickness.

\section{Particle migration under a transverse force}

\label{migr_norm}

The theoretical model described above corresponds to an idealized situation of migration in an
unbounded shear flow and without external forces acting.
In this Section we consider the channel flow, where the interactions with the walls
 should be taken into account. Besides, the particle may also experience an extra transverse force $F_{ex}\mathbf{e}%
_{y}$ induced by external fields (gravitational, electric, magnetic) or an additional
hydrodynamic force, such as, for example, the Dean force (in curved channels) or the so-called neutrally buoyant lift force
$F_{l}^{nb}$.

The force $F_{l}^{nb}$ is
usually evaluated numerically, assuming that the particle is free to rotate
and move in the $x$-direction, but is fixed in the transverse direction\cite%
{nakagawa2015,liu2015,lashgari2017}. Then, inertial migration of a particle to its
equilibrium position is simulated by balancing the lift force $%
F_{l}^{nb}$ and the transverse drag, i.e. the Saffman lift force $F_{l}^{Sa}$
is neglected although it can be significant at finite $\Re _{G}.$

%Indeed, there is some experimental \cite{zhou2020mapping} and numerical \cite{shao2008inertial} evidence of particle migration in the direction opposite to the neutrally-buoyant lift force at $\Re_G\sim20$.
%Some simulations of free particle motion in circular Poiseuille flow using the fictitious domain method\cite{shao2008inertial} demonstrated oscillatory motion of particles $\Re _{G}\simeq 20$, which means that their migration is not stationary.

Let us now generalize our analysis of instability to the case of a finite
transverse force $F_{ex}(y_{p})$ which depends only on the particle position
$y_{p}$, but not on its migration velocity $V_{y}$. We make an additional assumption that the
Reynolds number $\Re _{V}$ is small and the force changes slowly during
particle migration, i.e. the characteristic migration time $H/\left\vert
V_{y}\right\vert $ is large compared to the hydrodynamic time scale $G^{-1}.$
Here $H$ is the characteristic length scale for the change of $F_{ex}$
(that is usually the channel width). This is justified, i.e.  the ratio of the
two time scales is large, provided particles are small
\begin{equation*}
\frac{GH}{\left\vert V_{y}\right\vert }=\frac{\Re _{G}}{\Re _{V}}\frac{H}{a}%
\gg 1\text{\quad as\quad }a/H\ll 1,\ \Re _{V}\ll \Re _{G}.
\end{equation*}
In this case the migration is quasi-steady so that
the acceleration term in Eq.~(\ref{mot_dl2}) can be ignored as in the
neutral stability regime. We further assume that the Saffman lift $F_{l}^{Sa}$ is
controlled by streamwise acceleration of the
particle due to its transverse motion, i.e. given by (\ref{saffman_vy}). For such a situation
Eq.(\ref{mot_dl2}) for the transverse momentum can be rewritten as%
\begin{equation}
0=-f_{y}\left( y_{p}\right) V_{y}+C_{l}\left( y_{p}\right) \mathrm{St}\left(
y_{p}\right) V_{y}+F_{ex}\left( y_{p}\right),  \label{mot_dln}
\end{equation}%
where the Stokes number is based on a local shear rate $G\left( y_{p}\right)$, and the coefficients $f_{y}\ $and $C_{l}$  should  depend on the location of the particle to account for   hydrodynamic interactions with the walls.

Eq.~(\ref%
{mot_dln}) allows one to obtain the migration velocity under a slowly varying
transverse force $F_{ex}$:
\begin{equation}
V_{y}=\frac{F_{ex}}{f_{y}-C_{l}\mathrm{St}}.  \label{vy}
\end{equation}%

We recall that  the migration velocity is usually determined by balancing $F_{ex}$ and $%
F_{dy}=f_{y}V_{y}$, so that $V_{y0}=F_{ex}/f_{y}.$  Since for neutrally
buoyant particles it is traditionally assumed that $f_{y}=1,$ the migration velocity is simply $%
V_{y0}=F_{l}^{nb}.$\cite{nakagawa2015,liu2015,lashgari2017} Our results, however, show that due to the effect of the Saffman
lift force $V_{y}$ significantly deviates from $V_{y0}$, especially when the denominator in (\ref{vy}) is small.

If we consider a small perturbation $\delta V_{y}$ of the quasi-steady velocity \eqref{vy}, one can find that transverse motion becomes unstable when $C_{l}\left( y_{p}\right)\mathrm{St}\left( y_{p}\right)-f_{y}\left( y_{p}\right)> 0$. Therefore,
we recover the stability criterion \eqref{st_cr} for the force-free case, but now it involves functions of $y_{p}$. This suggests that the motion can be unstable only in some parts of the channel.

Eq.(\ref{vy}) for the migration velocity can be rewritten as
\begin{equation}
V_{y}=\frac{F_{ex}}{f_{y}\left( 1-\Re _{G}/\Re _{cr}\right) },  \label{vy1}
\end{equation}%
which includes $\Re _{cr}$. The last equation should be used for interpreting experimental data on
migration velocity. One can also conclude that the application of a steady state model to calculate the lift force from data obtained at finite $\Re _{G}$ can strongly overestimate the result, and would also lead to
incorrect scaling relationships.

\section{Simulation method}

\label{s4}

To simulate the flow we use a 3D implementation of the lattice Boltzmann
method (LBM) with a 19 velocity, single relaxation time scheme and the
Batnagar Gross Krook (BGK) collision operator~\cite%
{benzi_lattice_1992,kunert2010random}. Particles are discretized
on the fluid lattice and implemented as moving no-slip boundaries following
Ladd~\cite{LaddVerberg2001}.
The relaxation time of the BKG collision
operator is fixed to unity leading to a kinematic viscosity of $\nu =1/6$. Here and below the
variables are given  in simulation units. In addition, we set the fluid density $\rho=1$.  Further implementation
details are provided in our previous publications\cite%
{janoschek2010b,kunert2010random,janoschek2014,Dubov14,asmolov2018,nizkaya2020}%
.

 The size of the computational domain in most simulations is $%
(N_{x},N_{y},N_{z})=(200,161,100)$. We used spherical particles with radius $%
a=8$, which provides $H/a \simeq 20$, and prolate spheroids with equatorial radius $a=4$ and polar radius $b=8$.
%Fixed velocity boundaries are implemented at the top and bottom channel walls using mid-grid bounce-back boundary conditions and all remaining boundaries are periodic.
To generate a shear flow we implement impermeable no-slip walls moving with opposite velocities ($V_{w}$ at the top wall and $%
-V_{w}$ at the bottom wall)\cite{HechtHarting2010} and impose periodic
boundary conditions in the other two directions. The generated shear rate in
simulation units is $G = 2V_{w}/(N_{y}-1)$.

To search the unstable regimes for the particle equilibrium position at the channel  mid-plane we vary $\Re _{G}$ in the range from $0.25$ to $2$ and the particle density $\rho _{p}$ in the range from $15$ to $200$ to obtain different values of the Stokes number $\mathrm{St}$. We set initial rotational velocity $\boldsymbol{\omega }_{z}=-G/2$
and translational velocity $\mathbf{V}=(0.1 G a,0,0)$, and fix
the transverse coordinate $y_{p}=0$ during $3 \times 10^4$ time steps, waiting for the
system to equilibrate. Then, we release the particle and track its position for
$\sim 10^{5}$ time steps. If the transverse coordinate $y_{p}$ grows with time exponentially the equilibrium is deemed unstable.

\section{Results and discussion}
\label{res}

It is of considerable interest to compare LBM simulation data with our analytical theory and to determine
the regimes of validity of the theoretical results. Here we present results of our simulations  together with specific calculations using
theoretical expressions.

\subsection{Lift-to-drag ratio and transverse drag}

According to Eq.\eqref{st_cr} the critical Stokes number that should give us the instability onset depends on the ratio of $C_l$ and $f_y$. We, therefore, begin with the investigation of these parameters.

\begin{figure}[tbp]
\centering
\includegraphics[width=1.0\columnwidth]{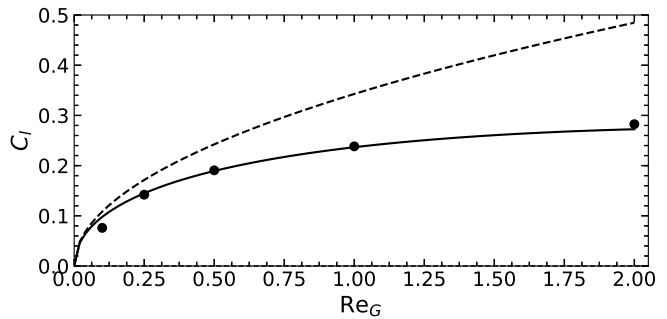}
\caption{Lift-to-drag ratio for a particle translating with a small slip
velocity (slip Reynolds number $\Re_V=0.05$) along the mid-plane. Circles show the simulation results. The
dashed and solid curves are calculations from Eq.\eqref{lift} and Eq.\eqref{cl_fit}. }
\label{fig:C_L}
\end{figure}

\begin{figure}[tbp]
\centering
\includegraphics[width=1.0\columnwidth]{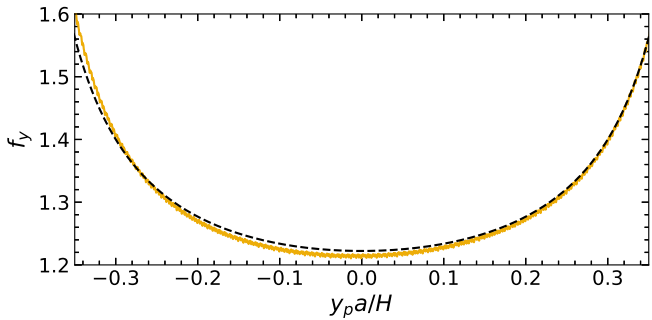}
\caption{Transverse drag for a particle translating with a small
slip velocity (slip Reynolds number $\Re_V=0.05$) across the channel in a
stagnant fluid. Solid curve shows simulation results,
dashed curve is calculated from Eq.~(\protect\ref{fy}).}
\label{fig:fy}
\end{figure}

To obtain the dependence of $C_{l}$ on $\Re _{G}$ we put the particle at the mid-plane of the channel and then move it in the $x$-direction with a velocity $V_{x}$ that corresponds to fixed $\Re_{V}=0.05$. By setting different $V_w$ we vary $\Re _{G}$ from 0.1 to 2. After equilibration, i.e. when the rotational
velocity of the particle becomes stationary, we measure the lift ($F_{l}^{Sa}$) and drag ($F_{dx}$) forces on the particle and average them over $10^{4}$ time steps. Then the lift-to-drag ratio is calculated as $C_{l}
=F_{l}^{Sa}/F_{dx}$. We remark that in these simulations we use our standard box that gives $H/a = 20$, but we have verified that the results for $C_{l}$ do not change if we set $H/a = 40$.

The computed lift-to-drag ratio $C_{l}
=F_{l}^{Sa}/F_{dx} $ as a function of $\Re _{G}$ is shown in Fig.~\ref{fig:C_L}. We see that on increasing $\Re _{G}$ the lift-to-drag ratio increases quickly when $\Re _{G}\ll 1$ and then shows a weak nonlinear growth.
The simulation data are compared with calculations from Eq.~\eqref{lift}, which is the Saffman formula derived for
small $\Re _{G}$. It can be seen that the Saffman formula fits well the simulation data obtained at $\Re _{G} \ll 1$, but strongly overestimates results at larger $\Re _{G}$. Also included in Fig.~\ref{fig:C_L} are calculations made using
\begin{equation}
C_{l}=0.343 \Re _{G}^{1/2}-0.106 \Re _{G}
\label{cl_fit}
\end{equation}%
obtained by fitting our data in the range $\Re _{G}\leq 2$. The first term here coincides with the Saffman lift $C_{l}^{Sa}$, and the second, linear in $\Re _{G}$, term is associated with a correction for finite $\Re _{G}$.

We now turn to the correction factor to the transverse drag $f_{y}$, which depends not only on $\Re _{G}$, but
also on $a/H$ and $y_p$ due to the wall effect. To obtain $f_y=F_y/V_y$ as a function of $y_p$ we place a particle at $y=0.4$, apply a small vertical force $F_y$, and then measure $V_y$ along the trajectory. In these simulations we fix $\Re _{G}=0$, i.e. perform measurements in a stagnant fluid since at finite $\Re _{G}$ it is difficult to distinguish between the transverse drag
and the lift force arising when the particle moves in the $y$-direction. We stress, however, that the effect of $H/a$ on $f_y$ is stronger than that of $\Re _{G}$. Consequently, the qualitative features of the $f_y$
curves at finite $\Re _{G}$ are the same, and the quantitative difference from the case of $\Re _{G}=0$ should be insignificant. Fig.~\ref{fig:fy} shows  $f_{y}$ plotted as a function of particle
position $y_p$ multiplied by $a/H = 0.05$. It has been earlier proposed that a sensible approximation for $f_{y}$ in the case of the channel can be simply a superposition of
single-wall contributions\cite{asmolov2018}

\begin{equation}
f_{y}=1+\dfrac{1}{H/2a-1+y_{p}}+\dfrac{1}{H/2a-1-y_{p}}.
\label{fy}
\end{equation}%
The calculations from Eq.\eqref{fy} are also shown in Fig.~\ref{fig:fy} and we see that the fit is quite good.
The function $f_y (y_p)$ takes its minimum value (of ca. 1.22 with our parameters) at the mid-plane, $y_p = 0$. Note that this exceeds $f_{y}=1$ corresponding to the Stokes drag in an unbounded flow. On approaching the walls $f_{y}$ increases, which implies that
the critical Stokes number $\mathrm{St}_{cr}$ given by Eq.\eqref{st_cr} also grows. Consequently,
inertial migration in the near-wall region
can remain stable even when the stability condition is violated in the central part of
the channel.

\subsection{Instability for spherical particles in Couette flow}

Next we examine the dependence of $\mathrm{St}_{cr}$ on $\Re_G$ and $\rho _{p}/\rho$. In these simulations particles are released at the mid-plane of the channel with a small initial velocity in the $x-$direction.

\begin{figure}[tbp]
\includegraphics[width=1.0\columnwidth]{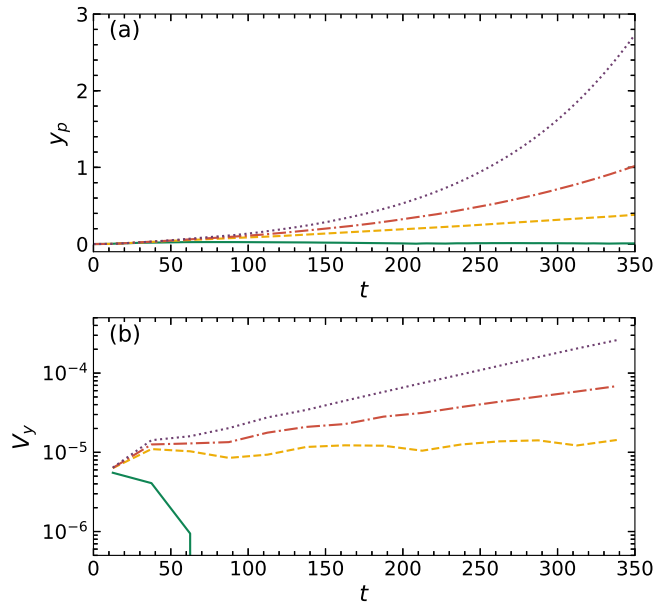}
\caption{Particle transverse positions (a) and magnitudes of their
transverse velocities (b) computed for the Couette flow using $H/a=20$, $\Re _{G}=0.5$
and $\mathrm{St}=5.6$ (solid), $7.8$ (dashed), $8.4$
(dashed-dotted), and $8.9$ (dotted). }
\label{fig:startup}
\end{figure}

Figure \ref{fig:startup} show the time dependence of particle trajectories and transverse velocities  obtained at $\Re _{G}=0.5$. These simulations are made using $\rho _{p}/\rho$ from 50 to 80, which corresponds to $\mathrm{St}$ varying from 5.6 to 8.9.
It can be seen that the particle with $\mathrm{St}=5.6$
remains at the mid-plane, but those with larger $\mathrm{St}$ accelerate in
the $y-$ direction, demonstrating the instability of the mid-plane
equilibrium (Fig.~\ref{fig:startup}(a)). In turn, the velocity $V_{y}$ reduces for the particle with $\mathrm{St}=5.6$, but augments exponentially with time if  $\mathrm{St}$ is larger (Fig.~\ref{fig:startup} (b)). The simulation data show that at large Stokes numbers the transverse velocity grows with their value.

\begin{figure}[tbp]
\centering
\includegraphics[width=1.0\columnwidth]{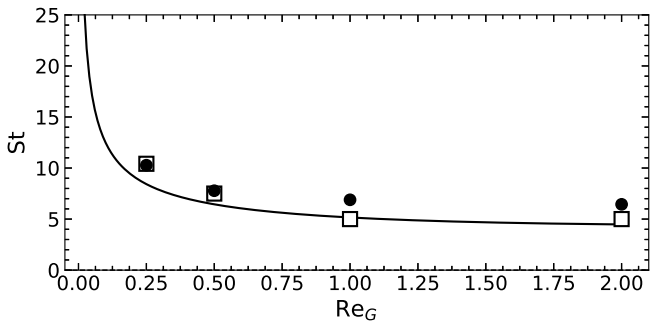}
\caption{Stability diagram for spherical particles in the Couette flow. Filled circles correspond to the onset of
instability obtained in simulations. The solid line is calculated from Eq.~\eqref{st_cr} using
$C_l$ given by Eq.~\eqref{cl_fit} and $f_{y}=1.22$. Open squares correspond to simulation data for spheroids.}
\label{fig:diagram}
\end{figure}

If similar analysis is made to a variety of simulations performed at different $\Re_{G}$ and  $\rho _{p}/\rho$, we can find $\mathrm{St}_{cr}$ that determines an onset of instability depending on these parameters. Fig.~\ref{fig:diagram} summarizes the simulation results (black circles) obtained for
$\Re_{G}$ from 0.25 to 2 and several density
ratios $\rho _{p}/\rho$ in the range from 15 to 200 in the $(\Re_{G}, \mathrm{St})$  plane. Note that the error
bars are smaller than the symbol size and, therefore, not shown. Also included is a theoretical, neutral equilibrium, curve calculated from Eq.\eqref{st_cr}. The calculations are made using $f_{y}=1.22$  and $C_l$ given by Eq.~\eqref{cl_fit}. An overall conclusion from this plot is that finite  $\Re _{G}$ dramatically reduce the value of $\mathrm{St}_{cr}$. We also conclude that the theory reproduces well the qualitative features of the neutral stability curve, although there is some quantitative discrepancy. The discrepancy is always in the direction of smaller $\mathrm{St}_{cr}$ than obtained in simulations, which is likely due to underestimated (obtained for $\Re _{G}=0$) $f_y$  used in theoretical calculations.

\begin{figure}[tbp]
\includegraphics[width=1.0\columnwidth]{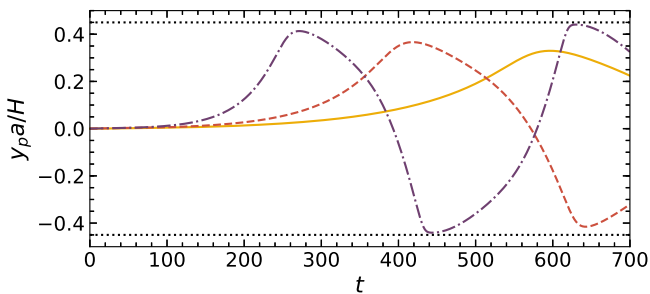}
\caption{Evolution of transverse positions in Couette flow for the particles
with $\Re _{G}=1$ and $\mathrm{St}=7.4$ (solid), $7.9$ (dashed) and $9$
(dashed-dotted). Dotted lines indicate contact with the walls.}
\label{fig:trajs_wall}
\end{figure}

We now fix $\Re _{G}=1$ and monitor the time evolution of $y_p$ at several supercritical $\mathrm{St}$. The results are plotted in Fig.~\ref{fig:trajs_wall}. It can be seen that after some interval of time particles discernibly deviate from their equilibrium position $y_p = 0$,  and that they move with the acceleration towards the wall indicating unstable equilibrium.  However, they slow down in the near-wall
region, and,  without making contact with the wall,  reverse the direction of  their motion. The particles then accelerate towards the opposite wall, etc. In other words, we observe the oscillations of particles between channel walls instead of their focusing at the mid-plane. Note that this oscillatory motion depends on $\mathrm{St}$. We see in Fig.~\ref{fig:trajs_wall} that the particles of larger $\mathrm{St}$ accelerate faster and oscillate with a smaller period, but larger amplitude. It is well seen that the extrema of $y_p$ become less pronounced and of smaller absolute value on decreasing $\mathrm{St} - \mathrm{St}_{cr}$. Clearly, the oscillations would disappear at $\mathrm{St} = \mathrm{St}_{cr}$ and smaller. These observations are, of course, very different from expected for
an unbounded shear flow, where the particle would accelerate until the lift force (which reduces with $\Re _{V}$) becomes equal to the transverse
drag. However, for our wall-bounded flow the local $\mathrm{St}_{cr}$
depends on $y_p$. Besides, in addition to the drag and the Saffman lift forces, the neutrally buoyant lift force $F_{l}^{nb}(y_p)$ (the force $F_{ex}$ in Eq.\eqref{mot_dln}) is acting on the particles. The latter does not depend on
$V_y$ and is caused by inertial hydrodynamic interactions with the walls\cite{Vass:Cox76}. Note that although this force is traditionally termed neutrally buoyant, it  would be the same for particles of any density. The local $\mathrm{St}_{cr}$ increases with the absolute value of $y_p$, i.e. on approaching the wall, since
$C_{l}$ decreases near the wall \cite{asmolov1990}, but $f_y$ is much
larger near the wall than in the central part of the channel (see Fig. \ref{fig:fy}). As a result, the particles retard near the wall and $V_y$ tends to zero, so does  the Saffman lift force given by \eqref{saffman_re}. One can speculate that particles commence the movement towards an opposite wall instead of immobilization
due to $F_{l}^{nb}$ that is directed away from
the wall\cite{Vass:Cox76}. They are pushed back to the mid-plane, but since the equilibrium there is unstable, continue to migrate until approach the wall.

\begin{figure}[tbp]
\includegraphics[width=1.0\columnwidth]{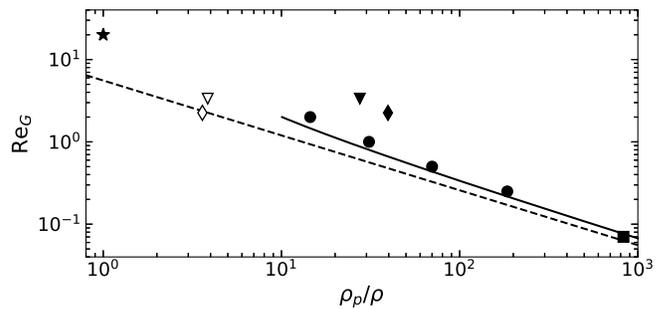}
\caption{The same as in Fig.~\ref{fig:diagram}, but plotted in the $(\rho_p/\rho,\Re_G)$
plane. Dashed line is calculated from Eq.~\eqref{re_cr}. The data inferred from earlier results for a pressure-driven flows are shown by filled triangles and diamonds (flat-parallel channels\cite{jebakumar2016lattice,zhang2016lattice}), the star (a circular tube\cite{shao2008inertial}), and the square (aerosol particles in a square
channel\cite{qian2020inertial}). The earlier data for steady-state trajectories\cite{jebakumar2016lattice,zhang2016lattice} are marked by the open triangle and diamond.}
\label{fig:scaling}
\end{figure}

Similar oscillatory trajectories have been found in simulations of neutrally buoyant\cite{shao2008inertial} and heavy particles\cite%
{jebakumar2016lattice,zhang2016lattice,qian2020inertial} migrating in pressure-driven
flows, but no attempt has been made to connect these results to the equilibrium instability.
In Fig.~\ref{fig:scaling} the data and the theoretical calculation are reproduced from Fig.~\ref{fig:diagram}, but plotted in the $(\rho _{p}/\rho,\Re_{G})$ plane and in a log-log scale. They are compared with the above mentioned simulation data obtained for pressure-driven flows and with
another calculation, made from Eq.\eqref{re_cr}. It can be seen that Eq.~\eqref{st_cr} provides quite good fit of our simulation data, but Eq.\eqref{re_cr} underestimates $\Re_{cr}$. We also conclude that earlier data for heavy particles in oscillatory regimes (filled symbols) always either fall into the instability region of a diagram or coincide with its onset. However, the data for steady-state trajectories (open symbols) fall into the stability region. Finally, we remark that the neutrally buoyant particles in an unstable equilibrium\cite{shao2008inertial} correspond to $\Re_G \simeq 20$, which is close to the values at which the Segre-Silberberg equilibrium position disappears, leaving only the inner annulus\cite{matas2003inertial,nakayama2019three}. Therefore, one can speculate that the particle inertia and the Saffman lift force may be important for interpreting this phenomenon too.

\subsection{Translational instability for spheroids}
\label{spheroid}

Our theory and above simulation results refer to spherical particles. Here we report some simulation data showing that our model could be suitable for spheroid particles too.

\begin{figure}[tbp]
\includegraphics[width=1.0\columnwidth]{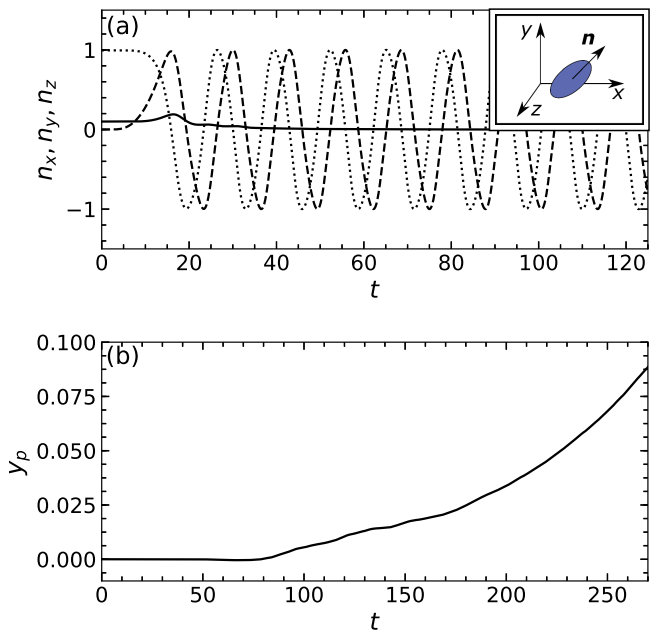}
\caption{Components of (a) the oritentation vector $n_x$ (dotted), $n_y$
(dashed), $n_z$ (solid) and (b) the transverse position of the prolate
spheroid in the supercritical regime at $\Re _{G}=0.5$ and $\mathrm{St}=7.8$.
}
\label{fig:trajs_ell}
\end{figure}

We investigate prolate spheroids of a polar radius $b$ and an equatorial radius $a$. In all simulations we use $H/b = 20$ and aspect ratio $b/a=2$. We define
\begin{equation}
\Re _{G}=\frac{Gb^{2}}{\nu },\quad \mathrm{St}=\frac{2\rho _{p}Ga^{2}}{9\mu }%
=\frac{2\rho _{p}}{9\rho }\left( \frac{a}{b}\right) ^{2}\Re _{G}.
\label{r_g2}
\end{equation}
and fix $\Re _{G}=0.5$ and $\mathrm{St}=7.8$. The spheroid is initially located at the mid-plane with some small inclination relative to $x-$axis and move with $V_x$. The time
evolution of the symmetry vector $\mathbf{n}=\left( n_{x},n_{y},n_{z}\right)$, which characterizes the particle orientation in the channel, is illustrated in Fig.~\ref{fig:trajs_ell}(a). We see that after some time a stable tumbling motion is established. This observation is in agreement with prior work\cite{huang2012rotation}. Simultaneously, the spheroid migrates in the $y-$direction with a growing with time velocity as seen in  Fig.~\ref{fig:trajs_ell}(b). This is exactly what we have observed for spheres (cf. Fig.~\ref{fig:startup}(a)). We have performed additional simulations using several
$\Re _{G}$ and $\mathrm{St}$. The results are included in Fig.~\ref{fig:diagram} and indicate that the onset of instability for our spheroid particles is very close to that for the spheres.

We have already mentioned some simulation studies\cite{qi2003rotational,huang2012rotation,rosen2015dynamical,rosen2016quantitative} of transitions between different rotational regimes of spheroids. However, these studies addressed  the case of $\Re_G \gg 1$ and used fixed coordinates, which implies that rotational and translational motions are decoupled. In  our work we have not fixed the spheroid position and   found that the instability can occur at much smaller Reynolds numbers, thanks to the Saffman lift force.

\section{Conclusion}

\label{con}

We have demonstrated theoretically that migration of inertial spherical particles in a shear flow becomes unstable, thanks to the Saffman lift force. It is shown that when their Stokes number exceeds the critical value, inertial particles migrate with an exponential acceleration. Lattice Boltzmann simulations of the critical Stokes numbers generally validate our analysis. Simulations also show that our simple theoretical model is also applied for 
prolate spheroids, and that  the lift-induced instability of
spheroid motion occurs approximately at the same Stokes numbers as for spheres. 

\begin{acknowledgments}
This work was supported by the Ministry of Science and Higher Education of the Russian Federation. We also acknowledge a partial financial support of by the Deutsche Forschungsgemeinschaft (DFG) under Project-ID 416229255 / SFB 1411 and FOR2688, grant HA4382/8-1.
\end{acknowledgments}

\section*{DATA AVAILABILITY}

The data that support the findings of this study are available
within the article.

\appendix

\section{Calculation of the lift force on a migrating particle}

In this Appendix, we derive the formula for the Saffman lift force on a
particle migrating in the neutral-stability regime.

\label{sec:lift} In his pioneering work Saffman\cite{saffman1965lift}
calculated a lift force on a sphere moving in unbounded shear flow $\mathbf{U%
}=y\mathbf{e}_{x}$ with a constant velocity $\mathbf{V}=V_{x}\mathbf{e}_{x}$
parallel to the flow, using the method of matched asymptotic expansions. In
the inertial coordinate system associated with the particle, $\mathbf{X}%
=\left( X,Y,Z\right) =\mathbf{x-x}_{p}$, the unperturbed flow reads
\begin{equation}
\mathbf{U}-\mathbf{V}=(Y-V_{s})\mathbf{e}_{x},  \label{und}
\end{equation}%
where $V_{s}=V_{x}-y_{p}$ is particle slip velocity.

The problem was solved in a strong shear limit, when the shear-based and
slip-based particle Reynolds numbers satisfy the condition
\begin{equation}
\Re _{V}\ll \Re _{G}^{1/2}\ll 1.  \label{str}
\end{equation}%
Condition (\ref{str}) means that the linear flow dominates over the slip
velocity in the outer region of the flow where $Y\sim \Re _{G}^{-1/2}$.
Therefore, far from the particle the unperturbed flow is $\mathbf{U}-\mathbf{%
V}\simeq Y\mathbf{e_{x}}$ and the disturbance induced by the particle
velocity $\mathbf{u}$ is governed by the Oseen-like equations,
\begin{equation}
\Re _{G}\left( Y\frac{\partial \mathbf{u}}{\partial X}+u_{y}\mathbf{e}%
_{x}\right) +\nabla p-\nabla ^{2}\mathbf{u}=6\pi \mathbf{F}_{p}\delta \left(
\mathbf{X}\right) .  \label{out}
\end{equation}%
Here, the terms in the brackets are the Oseen-like inertial terms, $\delta
\left( \mathbf{X}\right) $ is the delta-function, so that the particle
effect is approximated by the point force $\mathbf{F}_{p}$ exerted by the
particle on the fluid. For a particle moving with constant slip velocity $%
V_{s}$ this force is equal and opposite to the drag on the particle,\textbf{%
\ $\mathbf{F}$}$_{p}=-F_{dx}$\textbf{$\mathbf{e}_{x}$.} Therefore, the lift
force is proportional to the drag, and the ratio of the two forces is\cite%
{saffman1965lift}%
\begin{equation}
C_{l}^{Sa}=\frac{F_{l}^{Sa}}{F_{dx}}=0.343\Re _{G}^{1/2}\quad \text{for}%
\quad \Re _{G}\ll 1.  \label{lift}
\end{equation}%
Equations (\ref{out}) and (\ref{lift}) are usually written in terms of the
slip velocity $V_{s},$ since $F_{dx}=-V_{s}$ for the steady case at $\Re
_{V},\ \Re _{G}\ll 1.$

In our case the situation is different, since the particle accelerates in
the streamwise direction and migrates in the transverse direction. We
consider the flow using a non-inertial coordinate system translating with
the particle, $\mathbf{X}=\left( X,Y,Z\right) $.\cite{Maxey1983} The
unperturbed flow around the particle then reads
\begin{equation}
\mathbf{U-V}=Y\mathbf{e}_{x}-\left( V_{x}-y_{p}\right) \mathbf{e}_{x}-V_{y}%
\mathbf{e}_{y}.  \label{UV}
\end{equation}
For the neutral-stability regime, the streamwise component of the force $%
\mathbf{F}_{p}$ can be found by using Eqs.~(\ref{mot_dl1}), (\ref{dvt}),
while the transverse forces are balanced, and hence $\mathbf{F}_{p}=-\mathrm{%
St}V_{y}\mathbf{e}_{x}.$ Since the force is constant the disturbance flow is
steady. Assuming that $\Re _{V},\Re _{G}$ satisfy the condition (\ref{str}),
we can neglect the last two terms in (\ref{UV}) in the outer region.
Therefore, the disturbance velocity $\mathbf{u}$ in our case is governed by
the momentum equation similar to Eq.~\eqref{out}, with the drag force $F_{dx}
$ replaced by $\mathrm{St}V_{y}$. The lift force
on the particle in the neutral stability regime is then given by%
\begin{equation}
F_{l}^{Sa}=C_{l}^{Sa}\mathrm{St}V_{y}\quad \text{as}\quad \Re _{G}\ll 1.
\label{dvdt}
\end{equation}

\bibliography{lift_new}

\end{document}